\documentclass[prb,twocolumn]{revtex4-1}
\usepackage[table, dvipsnames]{xcolor}
\usepackage{graphicx}
\usepackage{amssymb}
\usepackage{dcolumn}
\usepackage{float}
\usepackage{bm}
\usepackage{multirow}
\usepackage{booktabs}
\usepackage[T1]{fontenc}
\usepackage[utf8]{inputenc}
\usepackage{lmodern}
\usepackage{color}
\usepackage{makecell}
\usepackage{tikz}
\usepackage{pgfplots}
\usepackage{mathtools}
\usepackage{physics}
\usepackage{amsmath}
\usepackage{setspace}
\usepackage{ragged2e}

\usepackage[colorlinks=true,linkcolor=red]{hyperref}

\newcommand{\mathleft}{\@fleqntrue\@mathmargin0pt}

\newcolumntype{M}[1]{>{\centering\arraybackslash}m{#1}}

\DeclareMathAlphabet{\bi}{OML}{cmm}{b}{it}

\def\be{\begin{equation}}
	\def\ee{\end{equation}}
\def\bearr{\begin{eqnarray}}
	\def\eearr{\end{eqnarray}}

\begin{document}
	
	\title{Unveiling the Chiral States in Multi-Weyl Semimetals through Magneto-Optical Spectroscopy}
	\bigskip
	\author{Sushmita Saha, Deepannita Das and Alestin Mawrie}
	\normalsize
	\affiliation{Department of Physics, Indian Institute of Technology Indore, Simrol, Indore-453552, India}
	\date{\today}
	\begin{abstract}
This study investigates the transport parameters in multi-Weyl semimetals, focusing on their magneto-optical properties and the role of chiral states. The tilting parameter is identified as a key factor in higher-order Weyl nodes, significantly influencing the magneto-optical response. We obtain a generic Landau-level expression for multi-Weyl semimetals, establishing a robust framework for analyzing their quantum transport properties. A comprehensive expression for the conductivity tensor components is presented, uncovering distinctive low-frequency peaks and other features shaped by the tilting parameter. Our findings reveal that the signatures of chiral states in the conductivity tensors become increasingly pronounced with the Weyl node order. Particularly, the tilting parameter is shown to impact Faraday rotation, at energies near the tilted Dirac cone energies. These results provide critical insights into the magneto-optical behavior of multi-Weyl semimetals and their potential for exploring topological phenomena.
	\end{abstract}
	
	\email{amawrie@iiti.ac.in}
	
	\maketitle
	
	\section{Introduction}
A multi-Weyl semimetal (mWSM)\cite{Fang,Liu,Dantas} is a unique quantum phase of matter, characterized by Weyl nodes in momentum space with topological charges exceeding one. Unlike conventional Weyl semimetals (WSMs),\cite{Armitage,Zyuzin,Yan} where Weyl nodes carry charges of \(\mathcal{C}=\pm 1\), mWSMs feature nodes with \(|\mathcal{C}| > 1\), resulting in enhanced Berry curvature and unconventional energy dispersions.\cite{WN_1,WN_2,WN_3,WN_4,WN_5}
The higher monopole charge (\(|\mathcal{C}| > 1\)) leads to nonlinear Landau level (LL) spacing,\cite{Liu} with unique scaling tied to anisotropic dispersion and Berry phase curvature near the nodes. The robustly chiral zeroth LLs of mWSMs host \(|\mathcal{C}|\) chiral states, enhancing topological robustness.\cite{Zeroth}
This amplifies Berry curvature effects, with intrinsic curvature and node separation influencing magneto-transport, magneto-optical phenomena, and Faraday and Kerr spectroscopy.\cite{Berry1,Berry2,Ma,Hosur,Orenstein,Dziom} The Faraday and Kerr spectroscopy illuminate key features like anomalous Hall conductivity, chiral edge states, and surface Fermi arcs, advancing topological photonics and non-reciprocal optics.\cite{Opti1,Opti2,Opti3,Opti4,Opti6,Opti7}

Using the generic LL expressions, we calculate the conductivity tensors, capturing the effects of higher-order monopole charges (\(|\mathcal{C}| > 1\)) and Berry curvature. This approach of detailing the conductivity tensor components offers a robust framework for quantum transport and magneto-optical responses, revealing key insights into mWSMs. Shaped by the multi-chiral nature of the zeroth LLs and the tilting parameter of Dirac cones,\cite{tilt1, tilt2, tilt3, tilt4, tilt5} these tensors exhibit pronounced signatures of chiral states. Notably, as the Weyl node order (\(|\mathcal{C}|\)) increases, the effects of the chiral states become more prominent, enhancing Berry curvature contributions and the topological response. 
The conductivity tensors unveil these signatures in Kerr and Faraday rotations, establishing a connection between magneto-optical responses and the underlying topology. A detailed investigation in a thin mWSM under linearly polarized light reveals the strong influence of Berry curvature on the polarization plane of reflected (Kerr) and transmitted (Faraday) light.
By varying the frequency and angle of incidence, we reveal how the anisotropic Berry curvature shapes these rotations, offering valuable insights for designing magneto-optical devices like optical isolators and polarization-sensitive photodetectors.
Our analysis reveals that the Faraday rotation is significantly influenced by the tilting parameter of the Dirac cones
, with pronounced effects observed at energies comparable to the tilted energy of these cones. The tilt parameter not only shapes the magneto-optical behavior but also offers a mechanism to reveal robust chiral states, as evidenced by distinct peaks in the Faraday rotation. These findings underscore the critical role of the tilting parameter in tuning the optical response and provide a deeper understanding of the magneto-optical behavior of mWSMs. Furthermore, they highlight the potential of these materials for exploring and utilizing topological phenomena, advancing theoretical understanding, and enabling applications in non-reciprocal optical devices, topological photonics, and quantum technology.

The paper is further structured as follows: Section \ref{Hamil_LL} discusses the Hamiltonian, LLs, and robustly chiral zeroth states in mWSMs. Section \ref{Opt_Ten} calculates the optical conductivity tensor, revealing light-matter interaction with the anisotropic Berry curvature field. Section \ref{fresnel} examines Kerr and Faraday rotations influenced by the material’s topological properties. Section \ref{C_S} concludes the study.


 \section{Model Hamiltonian and the Landau Levels}\label{Hamil_LL}
To describe the unique electronic properties of mWSMs, we use an effective low-energy Hamiltonian near a Weyl node:\cite{Berry1,Berry2,mW1,WN_1,mW3}  
\begin{equation}\label{Eq1}
\mathcal{H}_0(\mathbf{k}) = \lambda\left(k_-^m \sigma_+ + k_+^m \sigma_-\right) + \hbar \eta v_z \left(k_z + \eta {\mathcal{Q}}/{2}\right) \sigma_z,
\end{equation}
where \( k_\pm = k_x \pm i k_y \), \(\mathbf{k} = (k_x, k_y, k_z)\) is the crystal momentum, and the Weyl node is at \(\mathbf{k} = (0,0,\eta \mathcal{Q}/2)\). The parameters \( v_z \) and \(\lambda\) govern the velocity of the chiral states and Weyl-node strength, \(\sigma_{x,y,z}\) are Pauli matrices for the pseudo-spin, and \(\eta = \pm\) denotes chirality. The integer \( m>0 \) has a one-to-one relation with the topological number, \(\mathcal{C}\), thus distinguishes conventional (\(m = 1\) for \(\mathcal{C}=\pm 1\)) from multi-Weyl nodes (\(m \geq 2\) for \(|\mathcal{C}| \geq 2\)). 
In realistic crystal structures with high-symmetry points such as those with \( C_2 \) or \( C_3 \) rotational symmetry Weyl points can exhibit additional anisotropy due to tilt effects, which profoundly influence the physical properties of these materials. To account for this, we introduce a tilt Hamiltonian term, which represents both a linear tilt along the \( k_z \)-axis and a quadratic tilt in the \( k_x \)-\( k_y \) plane. \cite{tilt1,tilt2,tilt3,tilt4,tilt5,Armitage,Fang,Goerbig}
\begin{equation}\label{Eq2}
\mathcal{H}_{\rm tilt} = w_\parallel k_\parallel^2 + w_z k_z.
\end{equation}
The linear tilt term, \( w_z k_z\), effectively shifts the energy in the \( k_z \) direction, resulting in a “tilted cone” structure. This shift breaks isotropy and gives rise to unusual responses to external fields, impacting the transport and optical properties along the \( k_z \)-axis. Meanwhile, the quadratic tilt term, \( w_\parallel k_\parallel^2\), with \(k_\parallel=\sqrt{k_x^2+k_y^2}\), introduces a parabolic shift in the \( k_x \)-\( k_y \) plane, modifying the energy dispersion according to the material’s rotational symmetry constraints. Together, these tilt terms extend the effective model for mWSMs, allowing us to realistically account for the symmetry and structural constraints around Weyl points and enhancing our understanding of anisotropic behavior, such as direction-dependent transport and optical responses that emerge from these tilt effects.
The typical system parameters are listed in the table [\ref{Table1}].
\begin{table}[!ht]
    \centering
    \caption{Typical system parameters for mWSMs}
    \label{Table1}
    \begin{tabular}{
        >{\centering\arraybackslash}p{1.3cm} 
        >{\centering\arraybackslash}p{1.8cm} 
        >{\centering\arraybackslash}p{1.5cm} 
        >{\centering\arraybackslash}p{1.8cm} 
        >{\centering\arraybackslash}p{1.6cm} }
        \toprule
\hline        
        Winding Number & $v_z$ (eV$\mathring{\rm A}$) & $w_z$ (eV$\mathring{\rm A}$) & $w_{\vert \vert}$ (eV$\mathring{\rm A}^2$) & $\lambda$ (eV$\mathring{\rm A}^m$) \\ 
        \midrule
\hline
        \( m = 1 \) & 1 & 0.5 \(v_z\) & 0.05 & 0.4 \\
        \( m = 2 \) & 1.5 & \( 0.5\) \(v_z\)  & 0.05 & 0.2 \\
        \( m = 3 \) & 2 & 0.5 \(v_z\) & 0.02 & 0.08 \\
        \( m = 4 \) & 2.5 & 0.3 \(v_z\) & 0.02 & 0.04 \\
        \hline
        \bottomrule
    \end{tabular}
\end{table}
\subsection{The Generic Landau Levels}
We now calculate the LLs for the Hamiltonian $\mathcal{H}_{0}+\mathcal{H}_{\rm tilt}$, 
in the presence of a magnetic field \(\mathbf{B} = B_0 \hat{z}\). 
Theoretically, we introduce the magnetic field via the minimal coupling substitution: \(
\mathbf{k} \to \bm{\Pi} = \mathbf{p} + {e}\mathbf{A}/{\hbar},
\)
where \(\mathbf{A}\) is the vector potential satisfying \(\mathbf{B} = \nabla \times \mathbf{A}\). Using the Landau gauge, \(\mathbf{A} = (0, B_0 x, 0)\), the momentum components transform as:
\(
k_x \to \Pi_x = -i\hbar \partial_x, \quad k_y \to \Pi_y = -i\hbar \partial_y + {eB_0 x}/{\hbar}, \quad k_z \to k_z \text{ (unchanged)}.
\)
The in-plane motion is quantized by defining the ladder operators:
\[
a = \frac{l_B}{\sqrt{2}\hbar}\left(\Pi_x - {\Pi_y}\right), \quad a^\dagger = \frac{l_B}{\sqrt{2}\hbar}\left(\Pi_x + {\Pi_y}\right),
\]
where \(l_B = \sqrt{\hbar / eB_0}\) is the magnetic length. These operators satisfy the commutation relation \([a, a^\dagger] = 1\). The ladder operators \(a\) and \(a^\dagger\), act on the harmonic oscillator eigenstates \(|n\rangle\), such that
\(
a^\dagger a |n\rangle = n |n\rangle.
\)
On using these preliminary results
, we have
\begin{eqnarray}
    \mathcal{H}&=&\left( k_z + \eta \frac{\mathcal{Q}}{2} \right) (w_z \sigma_0 + \eta v_z \sigma_z) 
+ W_\parallel\left( 2 \hat{a}^\dagger \hat{a} + 1 \right) \sigma_0 
\nonumber\\&&+ \tilde{\lambda}_m
\begin{pmatrix}
    0 & \hat{a}^m \\
    (\hat{a}^\dagger)^m & 0 
\end{pmatrix},
\end{eqnarray}
where $W_\parallel = {w_\parallel }/{\ell_B^2}$ and $\tilde{\lambda}_m = \lambda ({\sqrt{2}}/{\ell_B})^m$.
\begin{figure}[http!]		\includegraphics[width=90.5mm,height=70.5mm]{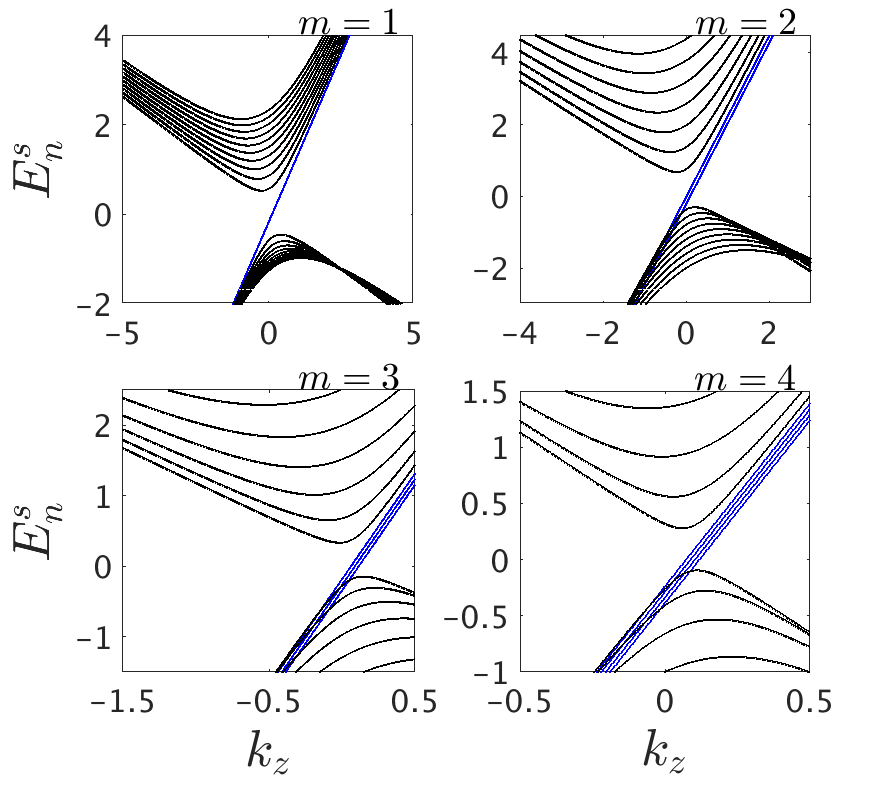}	
\caption{The LL dispersion (in eV) about the \(k_z\) direction (in units of 1/\(\mathring{\rm A})\) highlighting the emergence of robust $m$ chiral states. Parameters are taken from Table. [\ref{Table1}] with $\mathcal{Q}=0.25$ (in units of 1/\(\mathring{\rm A})\) and $\eta=-1$.}
		\label{Fig1}
	\end{figure}
For LLs in the bulk ($n \ge m$), we chose a spinor $|n,s\rangle=\begin{pmatrix}
    \alpha_n^s |n-m\rangle &
    \beta_n^s |n\rangle\end{pmatrix}^\prime$, where ``$^\prime$'' denotes the transpose and \(|n\rangle\) the $n^{\rm th}$ harmonic oscillator state such that 
\begin{eqnarray*}
    &&\Big[\left( k_z + \eta \frac{\mathcal{Q}}{2} \right) (w_z \sigma_0 + \eta v_z \sigma_z) 
+ W_\parallel \left( 2 \hat{a}^\dagger \hat{a} + 1 \right) \sigma_0 
\nonumber\\&&+\tilde{\lambda}_m
\begin{pmatrix}
    0 & \hat{a}^m \\
    (\hat{a}^\dagger)^m & 0 
\end{pmatrix}-E_n^s\Big]\begin{pmatrix}
    \alpha_n^s |n-m\rangle \\
    \beta_n^s |n\rangle
\end{pmatrix}=0,
\end{eqnarray*}
giving the eigenvalues for \(n>m\), \begin{align}
  E_n^s = (2n-m+1) W_\parallel + w_z \left(k_z + \eta \frac{\mathcal{Q}}{2}\right) + s \Gamma_{n}^m(k_z)
\end{align}
where \(
\Gamma_{n}^m (k_z) = \sqrt{\left[-m W_\parallel + v_z \eta \left( k_z + \eta \frac{\mathcal{Q}}{2} \right) \right]^2 + {\tilde{\lambda}_m}^2\frac{ n!}{(n - m)!}}
\), and 
\begin{equation}
\left.
\begin{split}
&\alpha_n^s = \frac{1}{\sqrt{2}} \sqrt{1 - s \frac{m W_\parallel - v_z \eta \left(k_z + \eta \frac{\mathcal{Q}}{2}\right)}{\Gamma_{n}^m}}\\
&\beta_n^s = \frac{1}{\sqrt{2}} \sqrt{1 + s \frac{m W_\parallel - v_z \eta \left(k_z + \eta \frac{\mathcal{Q}}{2}\right)}{\Gamma_{n}^m}}
\end{split}
\right\},
\end{equation}
In mWSMs, the zeroth LLs remain robust and chiral, characterizing their topological nature. They arise for \(n<m\), with the spinor being $\begin{pmatrix}
    0 &|n\rangle
\end{pmatrix}^\prime$, and the eigenvalues
\begin{align}
E_n=\left(k_z+\eta \frac{\mathcal{Q}}{2}\right)(-\eta v_z+w_z)+(2n+1)W_\parallel.
\end{align}
The said Landau spectrum is shown in Fig [\ref{Fig1}]. The plot of LLs in mWSMs reveals unique features compared to conventional WSMs. In mWSMs, due to the higher topological charge (\(m > 1\)) of the Weyl nodes, the LLs can be seen to exhibit nonlinear spacing. This nonlinear spacing arises from the anisotropic dispersion and the altered Berry curvature near the Weyl nodes. The \(m\) number of zeroth LLs, which remains chiral, shows a multiplicity of chiral states, further enhancing the topological effects. The energy levels as a function of the Landau index display distinct features, with the nonlinear energy dispersion becoming more pronounced as the order of the Weyl nodes increases, offering a visual signature of the multi-Weyl nature in mWSMs.

\section{Optical Conductivity Tensor}\label{Opt_Ten}
The Kubo formalism for magneto-optical conductivity can be expressed mathematically in terms of the Kubo formula, which relates the current density to the applied electric field. For a system subjected to a weak external electric field \( \boldsymbol{ \varepsilon} \), the linear response of the current density \( j_p \) can be described by the following expression\cite{LR1}
\[
j_p(\omega) = \sum_{q} \sigma_{pq}(\omega) \varepsilon_q(\omega)
\]
where \( \sigma_{pq}(\omega) \) is the conductivity tensor, which depends on the frequency \( \omega \) of the applied field, and \( p, q\) denote the \((x,y,z)\)-spatial components.
In the case of magneto-optical conductivity, the Kubo formula can be specifically formulated as:
\[
\sigma_{pq}(\omega) = i\hbar\int_0^\infty dt \, e^{i\omega t} \langle j_p(t) j_q(0) \rangle
\]
where \( \langle j_p(t) j_q(0) \rangle \) is the time-ordered correlation function of the current densities, defined in the Heisenberg picture.
Specifically, the current density operator \( \hat{j}_p \) associated with the \( p \)-th spatial direction is given by \(
\hat{j}_p = {e}/{\hbar} {\partial \mathcal{H}}/{\partial k_p},
\)
where \( e \) is the electronic charge, \( \hbar \) is the reduced Planck’s constant, and \( k_p \) denotes the \( p \)-th component of the crystal momentum, respectively. The operator evolves according to the Heisenberg equation of motion $
\hat{j}_p(t) = e^{i\mathcal{H}t/\hbar} \hat{j}_p(0) e^{-i\mathcal{H}t/\hbar}.
$
The correlation function can be computed by expressing \( \hat{j}_p \) and \( \hat{j}_q \) in the eigenbasis of \( \mathcal{H} \), and integrating over \( \mathbf{k} \)-space
\begin{eqnarray}
\langle j_p(t) j_q(0) \rangle &=& \sum_{n,n^\prime,s,s^\prime}[f_{n,s}-f_{n^\prime,s^\prime}]e^{i(E_n^s-E_{n^\prime}^{s^\prime})t/\hbar}\nonumber\\&\times&j_p^{n,s\rightarrow {n^\prime s^\prime}}j_q^{{n^\prime s^\prime}\rightarrow n,s},
\end{eqnarray}
\begin{figure*}[http!]		\includegraphics[width=185.5mm,height=75.5mm]{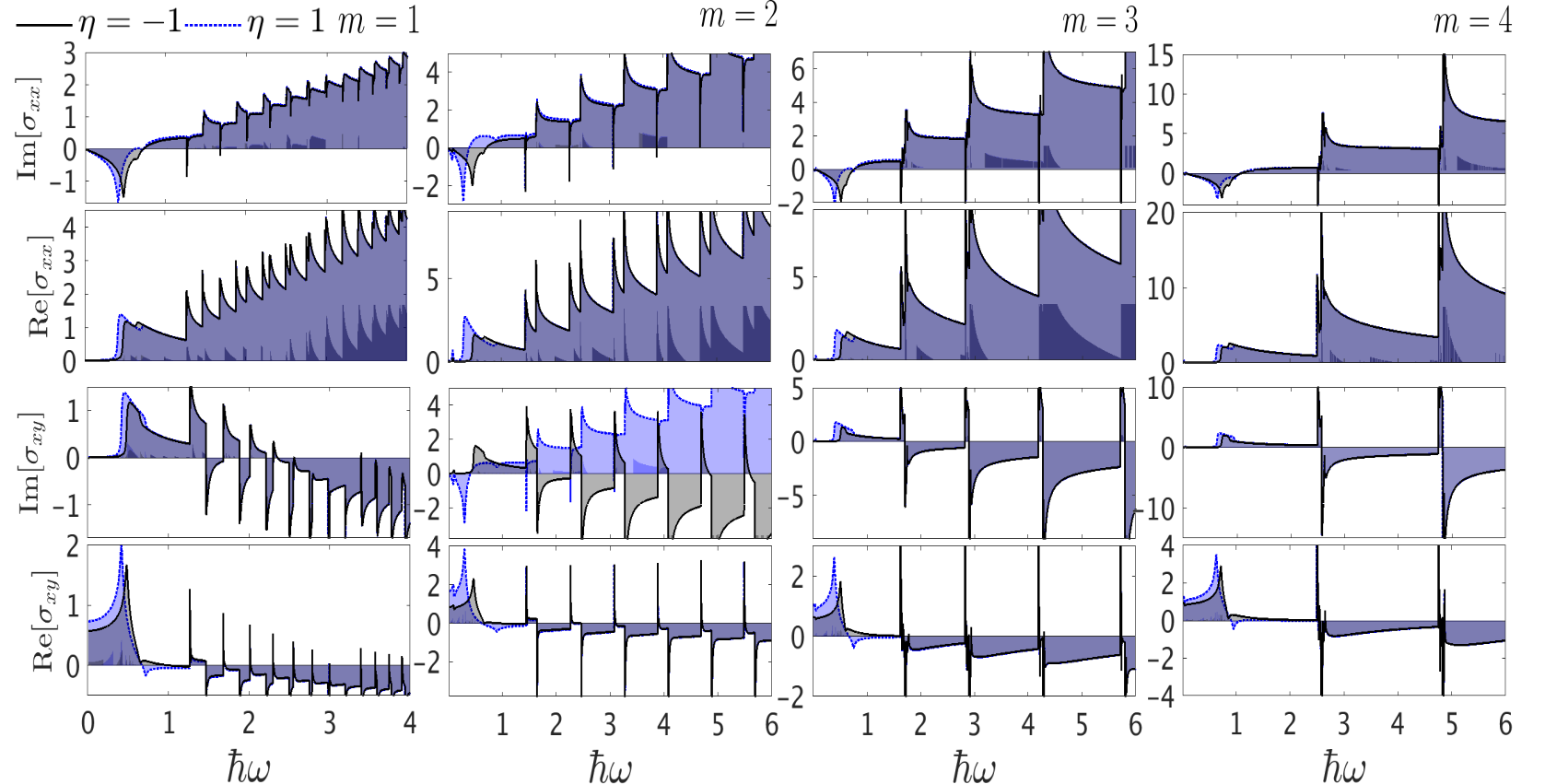}
\caption{The conductivity tensor $\sigma_L\;\& \; \sigma_H$ in units of $e^2/h$: From left to right m=1,2,3,4. Conductivity tensor components show enhanced chiral signatures with higher Weyl node orders and distinct Faraday rotation near tilted Dirac cone energies, highlighting the tilting parameter's role in mWSMs' magneto-optical behavior. The parameters are taken as $T=5$ K, the quantity `$0_+$' is set to $0_+=w_\parallel/5$ for $m=1,2$ and $0_+=w_\parallel/2$ for $m=3,4$ with $w_\parallel$ given in table \ref{Table1}, the magnetic length is normalised to $l_B=1$ and the rest of parameters are taken from Table [\ref{Table1}].}
		\label{Fig2}
	\end{figure*}
Here, $f_{n,s}$ is the Fermi Dirac function for the state \(|n,s\rangle\) and \(j_p^{n,s\rightarrow {n^\prime s^\prime}}\) is the matrix element for the current operator for a transition \(|n,s\rangle\rightarrow {|n^\prime s^\prime\rangle}\). Using all the above information we have the optical conductivity tensor components for transitions among the bulk states ($n \ge m$):
\begin{widetext}
\begin{equation}
\left.
\begin{split}
&\sigma_{yx} (\omega)= -\frac{e^2}{2 \pi \hbar} \sum_{n, s\ne s^\prime} \int_{-\infty}^\infty dk_z \, \zeta_n^m \Big[ \frac{1}{\hbar \omega + E_{n}^{s} - E_{n+1}^{s^\prime} + i0_+} - \frac{1}{\hbar \omega + E_{n+1}^{s^\prime} - E_{n,s} + i0_+} \Big] \frac{f_{n,s} - f_{n+1,s^\prime}}{E_{n}^s - E_{n+1}^{s^\prime}}\\
&\sigma_{xx} (\omega)= i\frac{e^2}{2 \pi \hbar} \sum_{n, s\ne s^\prime} \int_{-\infty}^\infty dk_z \, \zeta_n^m \Big[ \frac{1}{\hbar \omega + E_{n}^{s} - E_{n+1}^{s^\prime} + i0_+} + \frac{1}{\hbar \omega + E_{n+1}^{s^\prime} - E_{n,s} + {i0_+} }\Big] \frac{f_{n,s} - f_{n+1,s^\prime}}{E_{n}^s - E_{n+1}^{s^\prime}}
\end{split}
\right\},
\end{equation}
where \(
\zeta_n^m= ( \sqrt{2} W_\parallel \, \alpha_{n+1} \alpha_n \sqrt{n + 1 - m} + \frac{m}{\sqrt{2}} \, \tilde{\lambda}_m \, \beta_n \sqrt{\frac{n!}{(n + 1 - m)!}} \alpha_{n+1} + \sqrt{2} W_\parallel \, \beta_{n+1} \beta_n \sqrt{n + 1} )^2
\).
\end{widetext}
For transitions among the chiral states (\textit{viz} for $n < m$):
\begin{widetext}
\begin{equation}
\left.
\begin{split}
&\sigma_{yx} (\omega)= -\frac{e^2}{\pi \hbar} \sum_{n} W_\parallel^2 (n + 1) \int_{-\infty}^\infty dk_z \, \frac{f_{n} - f_{n+1}}{E_{n} - E_{n+1}}\Big[ \frac{1}{\hbar \omega + E_{n} - E_{n+1} + i0_+} - \frac{1}{\hbar \omega + E_{n+1} - E_{n} + i0_+} \Big]\\
&\sigma_{xx} (\omega)= \frac{i e^2}{\pi \hbar} \sum_{n} W_\parallel^2 (n + 1) \int_{-\infty}^\infty dk_z \, \frac{f_{n} - f_{n+1}}{E_{n} - E_{n+1}}\Big[ \frac{1}{\hbar \omega + E_{n} - E_{n+1} + i0_+} + \frac{1}{\hbar \omega + E_{n+1} - E_{n} + i0_+}\Big]  
\end{split}
\right\},
\end{equation}
And finally, for transition from the chiral state $(0\&|m - 1\rangle)^\prime$ to the state in the bulk ($|m,\pm\rangle$), we have
\begin{equation}
\left.
\begin{split}
&\sigma_{yx}(\omega) = -\frac{e^2}{2 \pi \hbar} \sum_{s} \int_{-\infty}^\infty dk_z \, \zeta_m \frac{f_{m, s} - f_{m - 1}}{E_{m}^s - E_{m - 1}}\left( \frac{1}{\hbar \omega + E_{m - 1} - E_{m}^s + i0_+} - \frac{1}{\hbar \omega + E_{m}^s - E_{m - 1} + i0_+} \right)\\
&\sigma_{xx} (\omega)= \frac{i e^2}{2 \pi \hbar} \sum_{s} \int_{-\infty}^\infty dk_z  \zeta_m \frac{f_{m, s} - f_{m - 1}}{E_{m}^s - E_{m - 1}}\left( \frac{1}{\hbar \omega + E_{m}^s - E_{m - 1} + i0_+} + \frac{1}{\hbar \omega + E_{m - 1} - E_{m}^s + i0_+} \right)
\end{split}
\right\},
\end{equation}
\end{widetext}
where \(\zeta_m=\left({m \sqrt{(m - 1)!/2}} \tilde{\lambda}_m \alpha_m + \sqrt{2} W_\parallel \beta_m \sqrt{m} \right)^2\).

We analyze these conductivity tensors in Fig [\ref{Fig2}]. The conductivity tensors shown in Fig [\ref{Fig2}] are profoundly influenced by the multi-chiral nature of the zeroth LL, which gives rise to distinct and pronounced signatures of chiral states. As the order of the Weyl nodes (\(m\)) increases, these chiral effects become more prominent, reflecting the deeper topological features of the system. The higher-order nodes contribute to an amplified Berry curvature, which significantly enhances the nontrivial topological response. This behavior not only underscores the role of topology in shaping the transport and optical properties but also highlights the potential of mWSMs for exploring complex quantum phenomena driven by their unique chiral nature.

\begin{figure}[http!]
		\includegraphics[width=85.5mm,height=35.5mm]{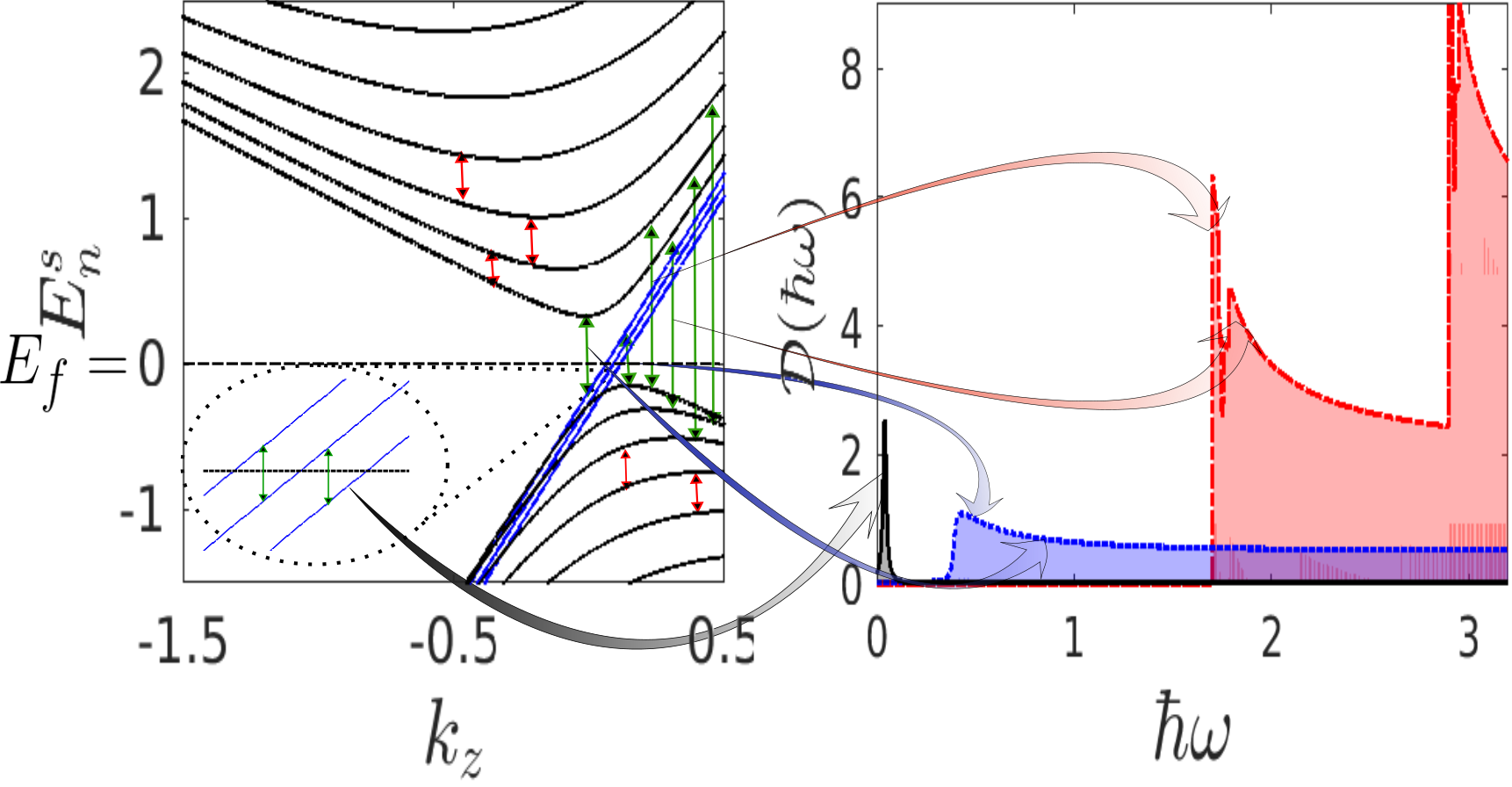}	
\caption{This figure maps the LLs spectrum (in eV) and the JDOS as a function of the frequency in the same eV units, illustrating transitions between chiral states and from chiral states to bulk states. For demonstration, we took $m=3$. }
		\label{Fig3}
	\end{figure} 
Characterizing the peaks in the magneto-optical conductivity tensors are the {joint density of states (JDOS)}, expressed as:
\[
\mathcal{D}(\omega) = \int_{-\infty}^\infty \sum_{n,n^\prime,s,s^\prime}\frac{dk_z}{2\pi} \, \delta(E_n^s(k_z) - E_{n^\prime}^{s^\prime}(k_z) - \hbar \omega).
\]
The Dirac delta function ensures energy conservation during the transition.
The JDOS plotted in Fig [\ref{Fig3}] is key in determining the spectral features of materials, such as absorption, emission, and scattering phenomena, by encapsulating the interplay between the energy landscape and transition probabilities. The figure illustrates the mapping of LLs spectra to the JDOS, highlighting the selection rules for optical transitions. In the left panel, the LLs spectrum is shown with \( k_z \) (wave vector along the \( z \)-direction) on the horizontal axis and \( E_n^s \) (LL energies) on the vertical axis.  The horizontal dashed line indicates the Fermi level, fixed at \( E_F = 0 \). Optical transitions are depicted using arrows: green arrows represent Pauli-allowed transitions, adhering to energy conservation, while red arrows denote Pauli-forbidden transitions, which are optically inactive. The right panel displays the JDOS as a function of photon energy \( \hbar\omega \), where peaks correspond to optical transitions between LLs. Large curvy arrows in the figure emphasize key transitions and their impact on the JDOS. The black arrow denotes transitions among chiral LLs, contributing unique features to the JDOS at low energies due to their linear dispersion. The blue arrow represents transitions from a chiral state \( |m-1\rangle \) to a bulk state \( |m\rangle \), indicating coupling between topological and bulk states, which influences the JDOS at intermediate energies. The red arrow highlights transitions among bulk LLs, dominating the JDOS at higher energies. Together, the figure provides a comprehensive depiction of how LLs physics, selection rules, and topological properties manifest in the optical response of systems like mWSMs as seen in Fig [\ref{Fig2}].

\begin{figure}[http!]
\includegraphics[width=85.5mm,height=30.5mm]{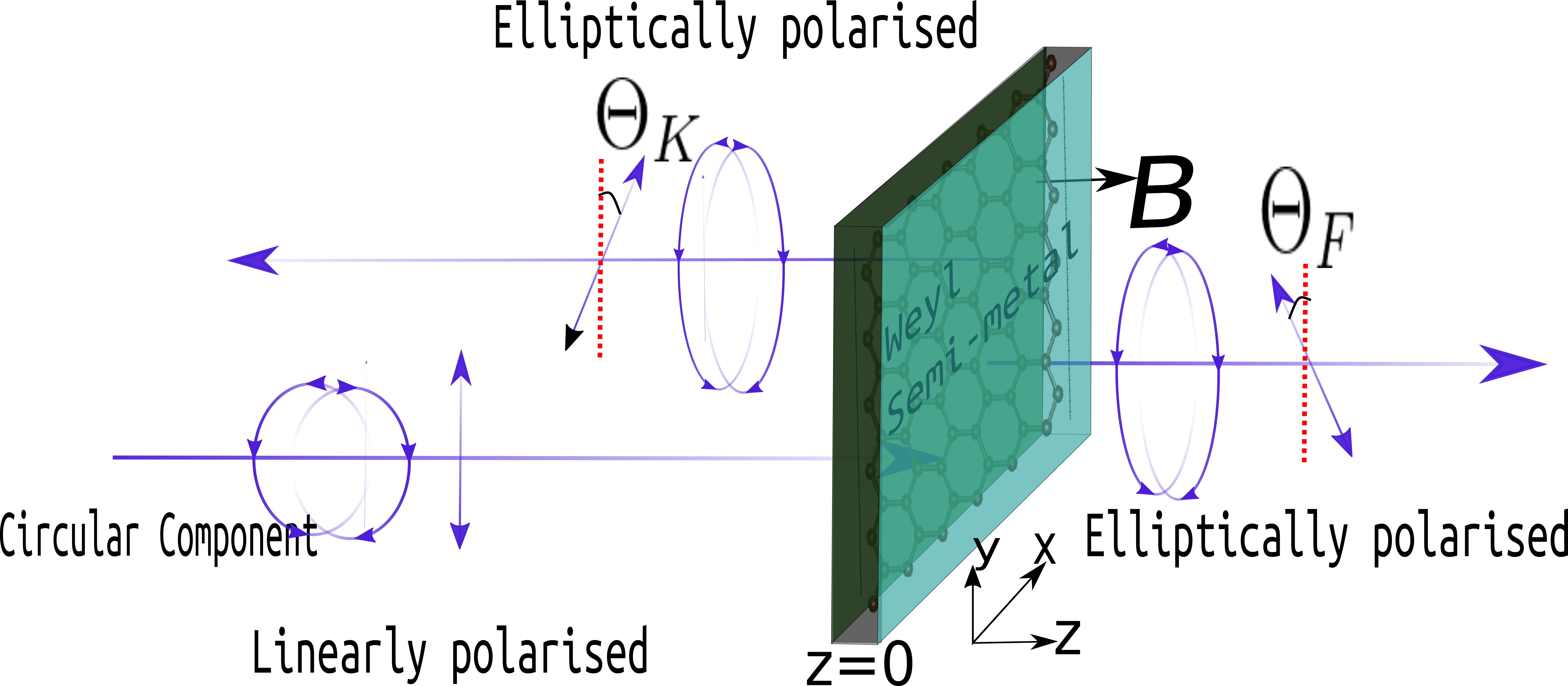}	
\caption{Schematic illustrating the Faraday and Kerr rotation in thin mWSMs, providing a framework to derive the Fresnel coefficients.}
		\label{Fig4}
	\end{figure}
\section{Fresnel Coefficients and the Faraday and Kerr spectrum}\label{fresnel}
\begin{figure}[b]
\includegraphics[width=50.5mm,height=30.5mm]{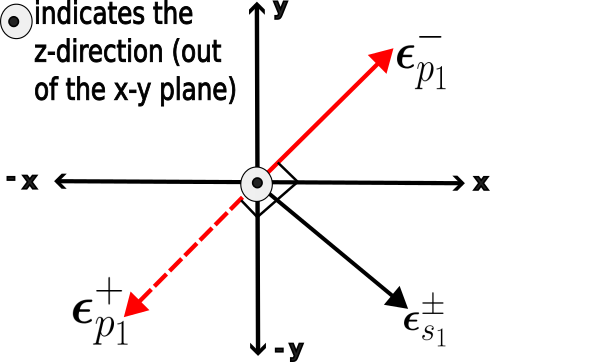}	
\caption{Depiction of the polarization vectors for s- and p-polarized waves in the medium 1. The red arrows indicate the polarization vectors $\epsilon_{p_1}^\pm$ when projected on the \((k_x-k_y)\) plane.}
		\label{Fig5}
   \end{figure}

For deriving the Fresnel coefficients needed to analyze the Faraday and Kerr spectrum, we refer to the schematic configuration shown in Fig. \ref{Fig4}, where an arbitrarily linearly polarized electromagnetic wave propagates through a vacuum before the incident on a mWSM thin slab at an angle $\theta_1$. Upon interaction with the surface of the semimetal, part of the wave reflects (leading to Kerr rotation, $\Theta_K$) while another part transmits into the medium (causing Faraday rotation, $\Theta_F$). The incident electric and magnetic fields can be expressed as:\cite{fresnel1}
\begin{equation}
\left.
\begin{split}
&\boldsymbol{\mathcal{E}}_I = \left[\mathcal{E}_I^s \boldsymbol{\epsilon}_{s_1}^+ + \mathcal{E}_I^p \boldsymbol{\epsilon}_{p_1}^+\right] e^{i(\kappa_{z,1}z + \mathbf{k} \cdot \mathbf{r} - \omega t)}\\
&\boldsymbol{\mathcal{H}}_I = \frac{1}{\mathcal{Z}_1} \left[\mathcal{E}_I^p \boldsymbol{\epsilon}_{s_1}^+ - \mathcal{E}_I^s \boldsymbol{\epsilon}_{p_1}^+\right] e^{i(\kappa_{z,1}z + \mathbf{k} \cdot \mathbf{r} - \omega t)}
\end{split}
\right\},
\end{equation}
Here, $\mathbf{k}$ is the in-plane wave vector, and $\boldsymbol{\epsilon}_s^\pm, \boldsymbol{\epsilon}_{p_{1,2}}^\pm$ represent the orthonormal polarization vectors in the respective medium (1,2). The superscript (+/-) represents the polarization vector of a particular wave defined in the medium (\(z<0/z>0\)), respectively. For better understanding, we sketch these vectors in Fig. [\ref{Fig5}]) for the vacuum medium and they can be written as\cite{fresnel1}
\begin{figure*}[t]
\includegraphics[width=180.5mm,height=75.5mm]{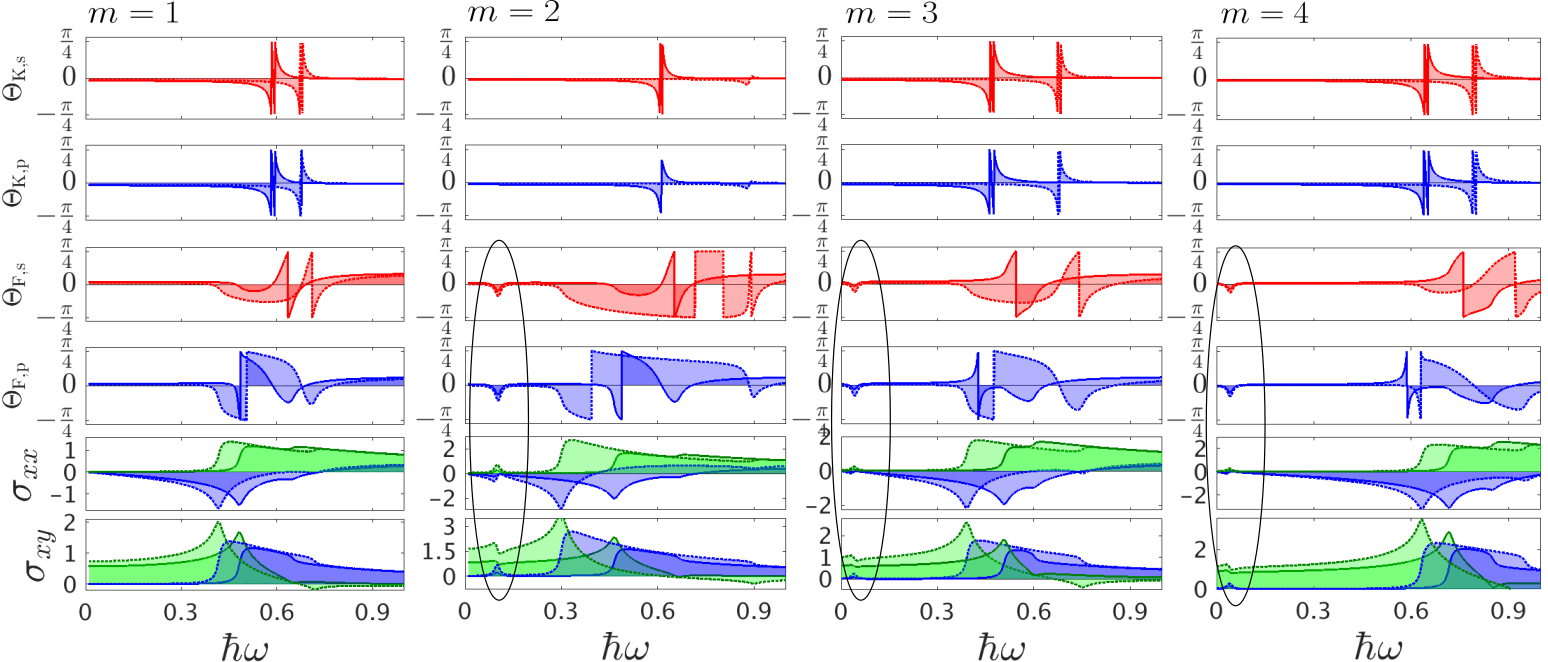}	
\caption{Magneto-optical Kerr ($\Theta_\text{K}$) and Faraday ($\Theta_\text{F}$) rotation angles, along with optical conductivity components ($\sigma_{xx}$ and $\sigma_{xy}$), as functions of photon energy ($\hbar \omega$) for mWSMs with topological charges $m = 1, 2, 3, 4$. The parameters to calculate the conductivity tensors is same as that in Fig. [\ref{Fig5}], and the angle of incident $\theta_1=\pi/4$ and the relative permitivity of the Weyl semi-metals is taken as $\epsilon_2=3$. The encircled regions highlight peaks associated with transitions between chiral states, indicating distinctive features in systems with higher topological charge.}
		\label{Fig6}
   \end{figure*}   
\begin{equation}
\left.
\begin{split}
&
    \boldsymbol{\epsilon}_{s_{1,2}}^\pm =\frac{k_y\hat{x} -k_x\hat{y}}{k} 
\\&\boldsymbol{\epsilon}_{p_{1,2}}^\pm= \frac{k}{k_0} \hat{z} \mp \frac{\kappa_{z_{1,2}}}{k_0}\frac{k_x\hat{x}+k_y\hat{y}}{k}
\end{split}
\right\},
\end{equation}
where $k_0 =\sqrt{k_z^2 +k^2}$, $\kappa_{z_n} =\kappa_n \cos{\theta_n}$ and $\kappa_n = \omega \sqrt{\mu_n \epsilon_n}$, with $\mu_n$, $\epsilon_n$ being the permeability and permitivity of the $n^{\rm th}$ medium. Similarly, the reflected and transmitted waves into the mWSMs, which are depicted in {Fig. [\ref{Fig4}]}, are given by:
 \begin{equation}
\left.
\begin{split}
&\boldsymbol{\mathcal{E}}_R = \left[\mathcal{E}_R^s \boldsymbol{\epsilon}_{s_1}^- + \mathcal{E}_R^p \boldsymbol{\epsilon}_{p_1}^-\right] e^{i(-\kappa_{z_1}z + \mathbf{k} \cdot \mathbf{r} - \omega t)}
\\
&\boldsymbol{\mathcal{E}}_T = \left[\mathcal{E}_T^s \boldsymbol{\epsilon}_{s_2}^+ + \mathcal{E}_T^p \boldsymbol{\epsilon}_{p_2}^+\right] e^{i(\kappa_{z_2}z + \mathbf{k} \cdot \mathbf{r} - \omega t)}
\end{split}
\right\},
\end{equation}

At the interface between the vacuum and the semimetal ($z=0$ when referring to {Fig. [\ref{Fig4}]}), the boundary conditions can be achieved by allowing the following conditions
\begin{enumerate}
    \item Continuity of the tangential component of the electric field:
\(
\mathbf{\hat{z}} \times \left(\boldsymbol{\mathcal{E}}_T - \boldsymbol{\mathcal{E}}_R - \boldsymbol{\mathcal{E}}_I\right) = 0.
\)
\item Discontinuity in the tangential magnetic field caused by the surface conductivity tensor $\boldsymbol{\sigma}$:
\(
\mathbf{\hat{z}} \times \left(\boldsymbol{\mathcal{H}}_T - \boldsymbol{\mathcal{H}}_R - \boldsymbol{\mathcal{H}}_I\right) = \boldsymbol{\sigma} \cdot \boldsymbol{\mathcal{E}}_T.
\)
\end{enumerate}
Using the above boundary conditions one can obtain the following relations:
\begin{equation}\label{continuity}
\left.
\begin{split}
&\mathcal{E}_T^s = \mathcal{E}_R^s + \mathcal{E}_I^s
\\
&\frac{\kappa_{z_2}}{\kappa_2} \mathcal{E}_T^p = \frac{\kappa_{z_1}}{\kappa_1} \left[ \mathcal{E}_I^p - \mathcal{E}_R^p \right]\\
& \frac{\mathcal{E}_R^p}{\mathcal{Z}_1} + \frac{\mathcal{E}_I^p}{\mathcal{Z}_1} = (\frac{1}{\mathcal{Z}_2} -\frac{\kappa_{z_2}}{\kappa_2} \sigma_L) \mathcal{E}_{T}^P - \sigma_{xy} \mathcal{E}_T^s\\
&  \frac{\kappa_{z_1}}{\mathcal{Z}_1 \kappa_1}  [\mathcal{E}_R^s - \mathcal{E}_I^s]  = \left[  \sigma_{T} -\frac{\kappa_{z_2}} {\kappa_2 \mathcal{Z}_2} \right] \mathcal{E}_T^s - \sigma_{H}\frac{\kappa_{z_2}}{\kappa_2} \mathcal{E}_T^p 
\end{split}
\right\},
\end{equation}
where $\mathcal{Z}_n=\sqrt{\mu_n/\epsilon_n}$ is the impedance of the $n^{\rm th}$ medium and ($\sigma_L,\sigma_T,\sigma_H$) are the conductivity tensor components ($\sigma_{xx},\sigma_{yy},\sigma_{xy}$), respectively. From the above Eq. [\ref{continuity}], one can determine the reflection and transmission coefficients defined as:
$ r^{i,j} ={\mathcal{E}_R^i}/{\mathcal{E}_I^j} \text{ and } t^{i,j} = {\mathcal{E}_T^i}/{\mathcal{E}_I^j},$ where $i,j$ can take all the possible combination of $s,p$ polarization to give us the following Fresnel coefficients\cite{fresnel1,fresnel2}
\begin{equation}
\left.
\begin{split}
&r_{ss} = -\frac{\Delta_+^L \Delta_-^T + \Delta^2}{\Delta_+^L \Delta_-^T + \Delta^2},  \text{ \;} r_{pp} = \frac{\Delta_-^L \Delta_+^T + \Lambda^2}{\Delta_+^L \Delta_+^T + \Lambda^2},
\\
&  t_{ss} = 2\frac{\kappa_{z_1}\Delta_+^L}{\Delta_+^L \Delta_+^T + \Lambda^2},  \text{ \;}  t_{pp} =2\sqrt{\frac{\varepsilon_{2}}{\varepsilon_{0}}}\frac{\kappa_{z_1}\Delta_{+}^{T}}{\Delta_{+}^{L}\Delta_{+}^{T}+\Lambda^{2}},\\
&     r_{ps} = -\frac{2\mathcal{Z}_0 \kappa_{z_1} \kappa_{z_2} \sigma_{xy}}{\Delta_+^L \Delta_+^T + \Lambda^2}, \text{ \;}   t_{ps} = \frac{\kappa_2 \kappa_{z_1}}{\kappa_1 \kappa_{z_2}} r_{sp},
\end{split}
\right\}
\end{equation}
where 
\begin{equation*}
\left.
\begin{split}
&\Delta_\pm^L = \epsilon_0\mathcal{Z}_0 n_2\omega\left(\pm \cos\theta_2 +n_2\cos\theta_1-\mathcal{Z}_0\cos\theta_1\cos\theta_2\sigma_L\right)
\\
&\Delta_\pm^T = \epsilon_0\mathcal{Z}_0 \omega\left(n_2 \cos\theta_2 \pm \cos\theta_1-\mathcal{Z}_0\sigma_T\right)\\
& \Lambda^2 = \mu_0^2 n_2 \omega^2\cos\theta_1\cos\theta_2 \sigma_{H}^2
\end{split}
\right\},
\end{equation*}
with $n_2=\sqrt{\epsilon_2/\epsilon_0}$ being the refractive index of the mWSMs. The standard definition of Faraday and Kerr angles are given by


\begin{equation}
\left.
\begin{split}
&    \boldsymbol{\Theta}_F^{s(p)} = \frac{1}{2} \tan^{-1} \left[ 2 \frac{\text{Re} \left[ \chi^{F, s(p)} \right]}{1 - |\chi^{F, s(p)}|^2} \right]
\\
&    \boldsymbol{\Theta}_K^{s(p)} = \frac{1}{2} \tan^{-1} \left[ 2 \frac{\text{Re} \left[ \chi^{K, s(p)} \right]}{1 - |\chi^{K, s(p)}|^2} \right]
\end{split}
\right\},
\end{equation}
where we define
\begin{equation}
\left.
\begin{split}
&\chi_F^p =\frac{t_{sp}}{t_{pp}}= -\mu_0\cos\theta_2\omega\sigma_H/\Delta_+^T\\
&\chi_F^s =\frac{t_{ps}}{t_{ss}}= -\mu_0 n_2\cos\theta_2\omega\sigma_H/\Delta_+^L\\
&\chi_K^s=\frac{r_{ps}}{r_{ss}}=\frac{2\mu_0 \sqrt{\mu_0\epsilon_0}n_2\cos\theta_1\cos\theta_2\omega^2\sigma_{H}}{\Delta_+^L\Delta_-^T+\Lambda^2}\\
&\chi_K^p=\frac{r_{ps}}{r_{ss}}=-\frac{2\mu_0 \sqrt{\mu_0\epsilon_0}n_2\cos\theta_1\cos\theta_2\omega^2\sigma_{H}}{\Delta_-^L\Delta_+^T+\Lambda^2}
\end{split}
\right\}.
\end{equation}
Fig. [\ref{Fig6}] presents the magneto-optical Kerr ($\Theta_\text{K}$) and Faraday ($\Theta_\text{F}$) rotation angles, alongside the real and imaginary parts of the optical conductivity tensor components ($\sigma_{xx}$ and $\sigma_{xy}$), as functions of photon energy ($\hbar \omega$) for mWSMs with topological charges $m = 1, 2, 3,$ and $4$. The analysis is focused on the low-frequency regime ($\hbar \omega \lesssim 0.5$), where the interplay among chiral states and that of the chiral state $|m-1\rangle
$ and the bulk state $|m,\pm\rangle$ produces distinct optical signatures. Notably, the encircled regions highlight peaks in the magneto-optical response, reflecting transitions between specific electronic states governed by the topological charge $m$. This approach aims to elucidate the significance of chiral states in shaping the magneto-optical properties of these materials.

In the low-frequency regime, two primary contributions dominate the magneto-optical response. The first is due to transitions among the chiral LLs, which emerge exclusively for $m > 1$. These transitions are absent for the trivial WSM case ($m = 1$), as it hosts only a single chiral state. The second contribution arises from transitions between the highest chiral state ($|m-1\rangle$) and the bulk states ($|m, \pm\rangle$). These transitions are prominent in the spectra and provide key insights into the interplay between topologically protected chiral modes and the bulk continuum. The figure reveals that as $m$ increases, the density of chiral states and the associated optical features become more pronounced, underscoring the role of topological charge in enhancing the chiral contributions to the magneto-optical response.

\section{Conclusions and Summary}\label{C_S}
In conclusion, this study offers a comprehensive exploration of the magneto-optical properties and quantum transport behavior of multi-Weyl semimetals, uncovering key insights into their unique topological and optical characteristics. By deriving a generic expression for the LLs, we establish a foundational framework for analyzing the quantum dynamics and energy spectrum of these materials, which are inherently influenced by their topological charge. The detailed expression for the components of the conductivity tensor further sheds light on the intricate interplay between chiral states and the bulk response, revealing their distinct contributions to the magneto-optical behavior.

Our findings emphasize the crucial role of chiral states, which become increasingly prominent as the order of the Weyl nodes grows. These states manifest as clear signatures in the magneto-optical conductivity and Faraday rotation spectra, particularly at energy scales comparable to the tilted energy of the Dirac cones. This underscores the significant influence of the tilting parameter in shaping the optical and transport responses of multi-Weyl semimetals. By capturing the interplay between topological charge, chiral states, and the tilting parameter, this work not only enhances the understanding of multi-Weyl semimetals but also highlights their potential as versatile platforms for investigating fundamental topological phenomena and developing novel magneto-optical applications.

	\textit{Acknowledgments}: This work is an outcome of
	the Research work carried out under the DST-INSPIRE project DST/INSPIRE/04/2019/000642, Government of India.

\end{document}